# Slow Light Propagation in a Linear-Response Three-Level Atomic Vapor

WENHAI JI, CHUNBAI WU and M. G. RAYMER

Oregon Center for Optics and Department of Physics, University of Oregon, Eugene, 97403 OR, USA; email: wji@uoregon.edu

(Dated: Dec. 2, 2006)

**Abstract.** We observe pulse delays of up to twenty times the input pulse duration when 200-ps laser pulses pass through a hot Rb 85 vapor cell. The pulse peak travels with a velocity equal to *c*/20, and the energy transmission is 5%. For pulses with linewidth greater than typical features in the atomic dispersion, pulse delay is predicted and observed for all center frequencies near resonance. Pulse advance is never observed. The measurements are in good agreement with a three-level linear-dispersion calculation. We are able to control the amount of delay by using a steady-state laser beam for optical pumping of the ground states prior to sending in the test pulse.

## 1. Introduction

Linear-response optical pulse propagation in a resonant dispersive medium is a classical problem. It has attracted interest since the early 1900's because of the possibility of observing fast and slow light behavior and its implications for Einstein's special theory of relativity. Brillouin in the 1930s initiated the theory of pulse propagation through a Lorentzian-absorber system [1]. In the anomalous dispersion region, the group velocity can be larger than the vacuum light speed, an effect (erroneously) called superluminal propagation. Brillouin [1], Icsevgi and Lamb [2] argued that such propagation does not violate relativistic causality for an infinite-support pulse, because the wave is already everywhere in space. The wave shape is just distorted through complex dispersion in the medium. For example, H. Tanaka, *et al.* [3] observed that the leading edge of the transmitted pulse never precedes that of the incident pulse. Thus, energy velocity and signal velocity [1, 4] were introduced to explain the operational meaning of pulse propagation and information exchange. For any type of pulse propagating through any type of system, the energy velocity (for steady-state) and signal velocity (for pulse) cannot exceed *c*, the vacuum speed of light.

Following these theoretical predictions, Chu and Wong [5] experimentally investigated the pulse propagation through a GaP:N crystal. Segard and Macke [6] put μs duration millimeter-wave pulses through a linear resonant molecular absorber, and recently Mikhailov, *et al.* [7] put optical pulses through a Rb 87 cell in the presence of two auxiliary light fields. Grischkowsky [8] observed slow pulse velocities in a Rb vapor with a pulsed magnetic field. His result agreed with an adiabatic following calculation. Sarkar, *et al.* [9] observed 200% fractional pulse delay when an 8 ps pulse passes through the GaAs quantum well at temperature of 20 K. Their tunability of optical delay was based on modification of group index by optically injecting free carriers. Wang, *et al.* [10] conducted a gain-assisted superluminal propagation experiment on the cesium D2 transition. The coherent gain-doublet structure was created for the Stokes Raman transition using a circular polarized optical pump and a Raman pump. The peak of the transmitted pulse appeared at the end of the cell 62 ns before the peak of the weak incident 3.7 μs probe pulse appeared at the entrance to the cell. However, the 0.8 ns pulse advance time was a small fraction of the 3.2 ns pulse width [3] and the 62 ns pulse advance time was a small fraction of the 3.7 μs pulse width [10]. As far as we are aware, the pulse advance is small compared with the pulse width in all 'superluminal' experiments reported so far. We shall return to this point later.

In some studies, the nonlinear optical process of self-induced transparency (SIT) has been accountable for the reduced group velocity of $2\pi$ pulses in a two-level system [11]. In addition, by applying a coherent pulse to a three-level $\Lambda$ system, electromagnetically induced transparency (EIT) spectral windows can be created within which there is steep slope of refractive index versus probe frequency and low absorption. EIT has a promising application in quantum memory [12-14] and quantum information processing because of its controllable pulse group velocity. However, EIT has limitations. The EIT windows should not be too wide, in order to have a steep slope of index. This limits the signal bandwidth, and the pulse width in EIT experiments is never narrower than ns. For our pulse slowing, a physical mechanism other than EIT and SIT is accountable—namely, simple linear dispersion arising from populated ground states.

In our experiment, we observe pulse peak delays equal to 20 times the duration of the input when a 200ps pulse passes through a Rb 85 cell. The corresponding peak traveling velocity of *c*/20 is slower than the early EIT results of *c*/13 [13]. H. Tanaka, *et al* [3] did a similar experiment. They observed pulse delay up to 10 times of pulse duration when a 3.2 ns pulse passed through hot natural Rb vapor cell. Superluminal propagation of a probe pulse was observed when its frequency is nearly resonant with the Rb 85 transition. Our study is distinct from that of Tanaka in that our pulses are shorter, making the simple concept of group velocity generally inapplicable. Also, we calculate the wave shape of the transmitted pulse and compare it with experimental observations. We predict that there exist oscillations in the output because of beating between two transmitted frequency components. We realized controllability of the pulse delay using optical pumping, which we can switch on and off rapidly.

**2. Experiment setup**

The experiment setup is shown in Figure. 1. The 80 MHz repetition rate pulses from a titanium-sapphire laser are amplified in a regenerative cavity, at a pulse repetition rate of 5 kHz. The pulse width (full width half maximum, FWHM) is 200 ps, measured by a fast photo diode and a sampling oscilloscope. The linewidth of the pulse is 2.3 GHz, measured by a scanning Fabry-Perot cavity. The center wavelength is tuned around the Rb 85 D1 transition region, centered near 794.78 nm. The wavelength is monitored by a Burleigh WA-1500 wavemeter. After the regenerative optical amplifier, which produces a pulse rate of 5 kHz, the beam is filtered out from the background emission by a grating and attenuated to average power 8 µW, and then collimated to a beam spot size of 1 mm.

The sample is a 75 mm long, isotopically pure Rb 85 vapor cell inside an oven, surrounded by a 0.04-inch thick mu-metal box. There is 4-5 torr of neon buffer gas in the cell that is used to increase the diffusion time of Rb atoms through the laser beam. The vapor cell is heated to 100 C, at which the atomic number density equals $N=4.8\times 10^{18}$ m$^{-3}$. A 1-ns rise-and-fall-time silicon photodiode D1 is used to measure the delay relative to a reference photo diode D2 that is located before the cell.

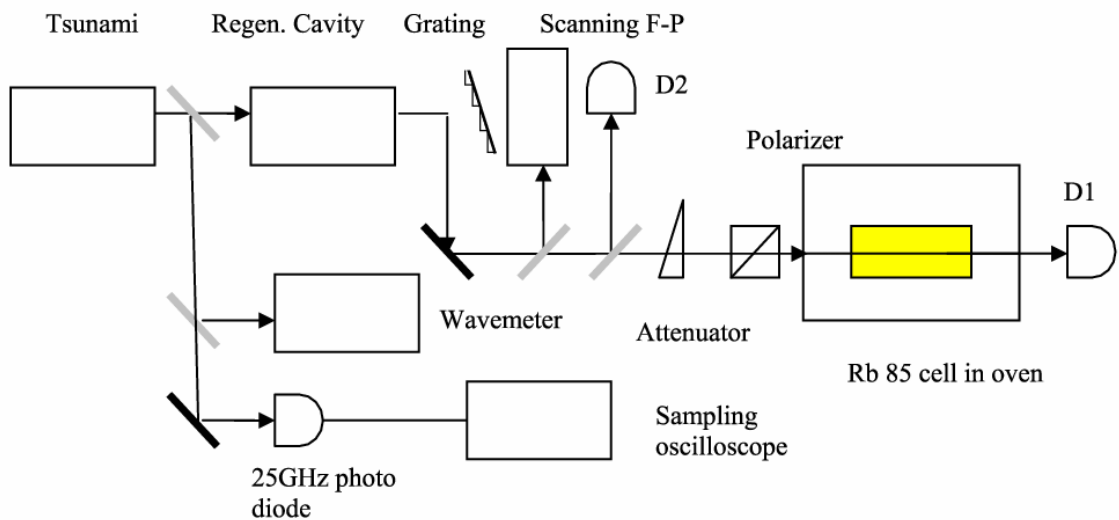

**Figure. 1.** Experiment setup: Laser pulses from a Tsunami titanium-sapphire laser are sent to the regenerative cavity to be amplified. Their wavelength and pulse mode locking stability are monitored by a wavemeter and fast photodiode correspondingly. After the pulses reflect from a grating, and pass through an attenuator and polarizer, they propagate through the hot Rb 85 vapor cell and reach the D1 detector.

In contrast, in the Tanaka experiment [3], the 1 cm Rb cell contains natural Rb (25% of Rb 87 and 75% of Rb 85), while the effective component that caused all pulse delay and advance in their and our experiment is Rb 85. They heated the vapor cell to 175 C to observe larger delay. They used slower detectors than ours. They used 3.2 ns pulses, 15 times our pulse duration. Their corresponding input pulse linewidth is narrower than the Doppler linewidth, whereas our pulse linewidth exceeds the Doppler linewidth. These differences enable us to observe qualitatively different phenomena. The frequency dependence of the pulse delay time in their experiment is different from our counterpart. For the same reason, we do not observe superluminal propagation.

## 3. Theoretical model

Since the laser power is weak, linear dispersion theory is applicable. The medium is a three-level atomic $\Lambda$ system (Figure. 2), described by a linear susceptibility for stationary atoms [15].

$$\chi(\nu) = \frac{N}{\hbar\varepsilon_0} \sum_{n,m} \frac{(\bar{\rho}_{mm} - \bar{\rho}_{nn})\mu_{mn}\mu_{nm}}{\omega_{nm} - \nu - i\gamma_{nm}} \qquad (1)$$

Here $N$ is the atomic number density, $\omega_{nm}$ is the transition angular frequency of level $n$ to level $m$ [16], $\nu$ is laser angular frequency, $\bar{\rho}_{mm}$, $\bar{\rho}_{nn}$ is the probability for an atom being in $m$, $n$ states. In Rb 85 D1 transition, $n, m = 1 - 4$ are corresponding to two ground states and two excited states indicated in Figure.2. We assume that initially there is no coherence between any states, and no population in any excited states. The degeneracy of the various levels implies the populations are $\bar{\rho}_{11} = 5/12$, $\bar{\rho}_{22} = 7/12$, $\bar{\rho}_{33} = \bar{\rho}_{44} = 0$. The decay rate $\gamma_{nm} = \gamma_r + \gamma_p$ includes half the radiative decay rate $\gamma_r = 2\pi \times 3$ MHz [17, 18] and the pure dephasing rate $\gamma_p$ caused by the Rb-Rb and Rb-Ne (buffer gas) collision. The Ne-induced dephasing rate is linearly proportional to the buffer gas pressure. The 4-5 Torr buffer gas pressure gives $\gamma_{nm} = 2\pi \times 25$ MHz [7, 19]. All these parameters are known from prior experiments. Thus, our simulations have no free parameters. We account for the contributions of atoms with different velocities in the Doppler profile (650MHz FWHM) to the overall atomic linear susceptibility by convolution of the single-atom susceptibility $\chi(\nu)$ with the Gaussian velocity distribution. The calculated real and imaginary parts of the susceptibility are given in Figure. 3. They are similar in shape to the standard EIT susceptibility [20]. Without any auxiliary optical field, Rb 85 naturally has an "EIT-like" structure. Unlike our method of obtaining the susceptibility from first principles, H. Tanaka, *et al.* [3] calculated the susceptibility from the measured absorption coefficient using the Kramers-Kronig relations. From our calculation, we are able to tell that the dominant factor that causes delay and superluminal effects is Rb 85. Different from the experiment by Wang, *et al.* [10] in which a gain doublet structure is created by two circular polarized optical fields, our experiment is based on an intrinsic absorption doublet structure.

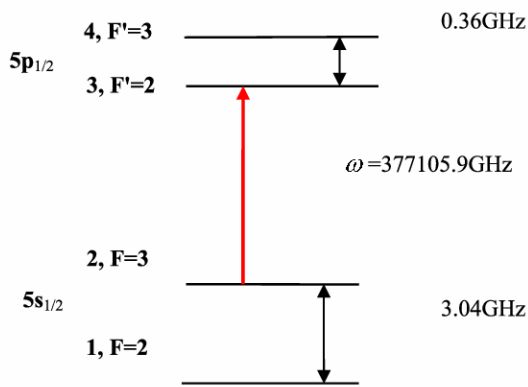

**Figure. 2.** Rb 85 D1 line. The hyperfine splitting in $5s_{1/2}$, $5p_{1/2}$ state is 3.04 GHz and 0.36 GHz respectively. The transition frequency between F=3 and F'=2 is 377105.9 GHz.

A pulse with sufficiently narrow linewidth propagates in the cell with group velocity that is related to the slope by:

$$v_g = \frac{c}{n + \nu \frac{\partial n}{\partial \nu}} = \frac{c}{n + \frac{1}{2}\nu \frac{\partial \mathrm{Re}\,\chi}{\partial \nu}} \quad (2)$$

Here $c$ is the speed of light in vacuum, $n$ is the refractive index, which is approximately 1.0 in Rb 85 vapor. If a pulse with narrow linewidth and center frequency in the middle of the low-absorption region passes the cell, slow light can be observed. In order to evaluate the 'quality' of the slowing down, we introduce a Figure of merit (*FoM*) defined as:

$$FoM = \frac{1}{|\frac{v_g}{c}\alpha|} \quad (3)$$

Here $v_g$ is the group velocity and $\alpha$ is the absorption coefficient. In Figure. 3, both linear susceptibility and *FoM* are given for quasi-monochromatic pulses. Based on the plot of *FoM*, for the low-group-velocity region, the best place for conducting the pulse delay experiment is the middle of the region, defined as zero frequency in the plots. The features of group velocity in the absorption region are due to hyperfine splitting in the excited state.

The measured product of input pulse duration (200 ps) and pulse bandwidth (2.3 GHz) is 0.46, which is close to the Fourier transform limit of 0.44 for an assumed Gaussian shaped pulse. For our probe pulse linewidth about 3 times broader than Doppler linewidth, we cannot calculate the group velocity using Eq. (2). For such a near transform-limited pulse, the output spectrum can be obtained by multiplying the input spectrum with a transfer function H($\nu$) in the frequency domain. The output pulse in the time domain can then be obtained from the output spectrum through a Fourier transform. The transfer function for an input field of $E(z,t) = A(z,t)\exp(i2\pi\nu t - ik_0 z)$ is [21].

$$H(v) = \exp\{-\frac{1}{2}\alpha(v)z - i[k(v) - k_0]z\}$$
$$= \exp\{-\frac{1}{2}k_0 z \operatorname{Im} \chi(v) - i\frac{1}{2}k_0 z \operatorname{Re} \chi(v)\} \quad (4)$$

Here $z$ is the length of the vapor cell, $k_0$ is the wave number in air, $v$ is the center laser frequency, $\alpha(v)$ is the absorption coefficient, $k(v)$ is the wave number in the medium with laser frequency $v$.

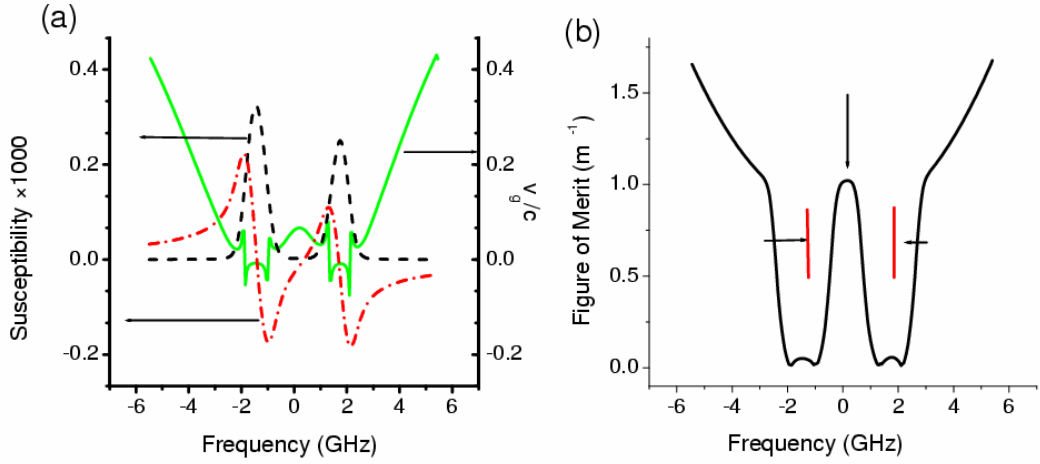

**Figure. 3.** Theoretical calculation of (a) real part (black dashed, scale adjusted) and imaginary part (red dash dotted, scale adjusted) of linear susceptibility, group velocity (green solid) vs. laser frequency and (b) Figure of merit vs. laser frequency for monochromatic light. Here and in following Figures, zero in the frequency axis corresponds to 377107.5 GHz, which is the middle frequency between the two resonant peaks 377106 and 377109 GHz. Here and in following Figures, experiment and simulation conditions are for pulse propagation in a 100 C, 75 mm long Rb 85 vapor cell, unless other specified.

**4. Experimental observation and discussion**

*4.1. Wave shape and spectrum*

As the 200-ps weak pulse propagates, the absorbing medium Rb 85 creates a hole in the pulse spectrum. The light beam develops temporal beats due to the superposition of the two resulting narrow spectral components that differ in central frequency. As the pulse propagates farther, the hole in the pulse spectrum becomes deeper (Figure. 4. a). This causes the two apparent side frequencies to become farther apart so that the temporal beats become more rapid (Figure. 5. a). The other way of understanding the pulse broadening is that the group velocities of the two frequency components are different. However, the detector with 1-ns rise-and-fall-time will not resolve the beat structure that oscillates in the 2-3 GHz range. Instead, only envelopes are recorded by the 200 MHz oscilloscope (Figure. 5. b). The recorded waves also do not have a clear, sharp front edge, an effect that is due to the pre-pulse from the low extinction ratio (around 10:1) regenerative cavity. In addition, the input spectrum, which includes a small percentage of background, is not perfectly Gaussian. The background light propagates in the vapor cell without slowing down or being absorbed. Thus, it makes the measured transmission appear higher, especially in the two resonant absorption regions.

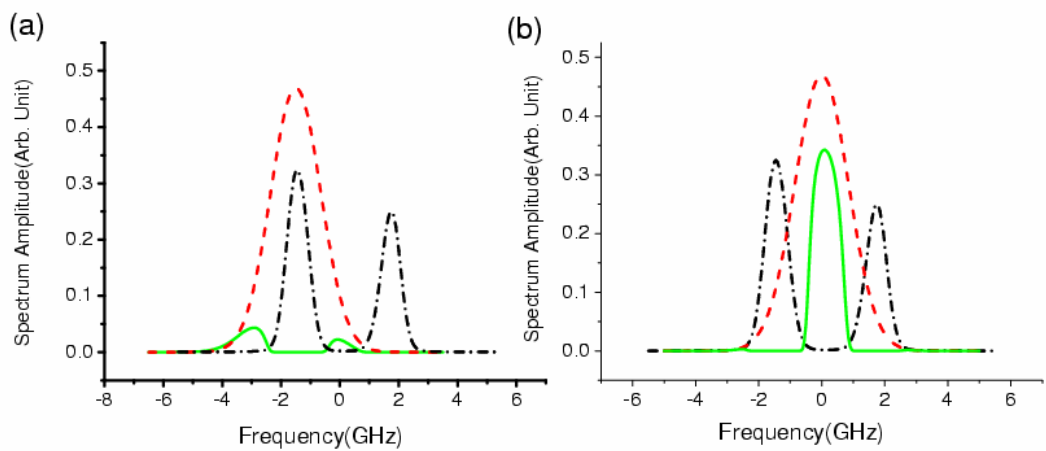

**Figure. 4.** The simulation result of both input and output spectrum at (a) 377106 GHz and (b) 377107.5 GHz. The Imaginary part of $\chi$ is also given (black dash dotted, scale adjusted) to indicate the two resonant absorption regions. The red dashed curve and green solid curve are the corresponding input and output spectra.

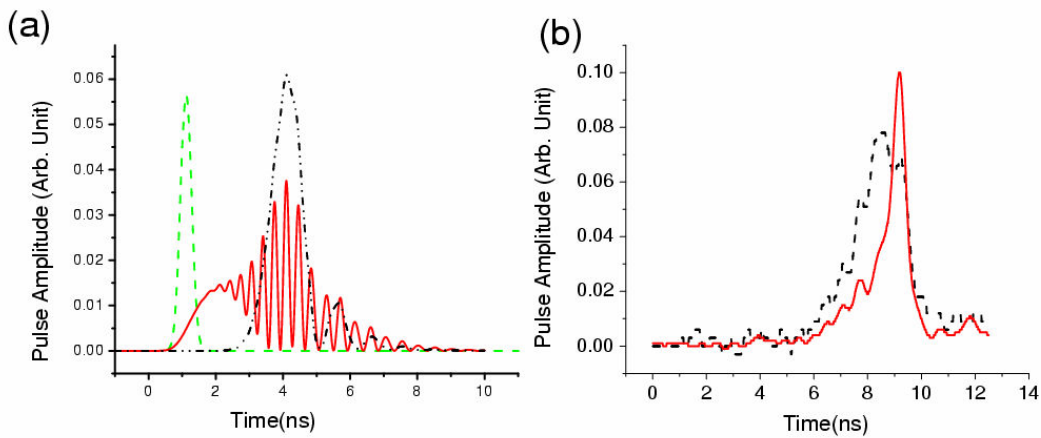

Figure. 5. Simulated and measured pulse shapes. (a) The wave shapes from simulation: the green dashed curve is the 200 ps input pulse, the black dash dot dotted curve and red solid curve are output shapes for center frequency at 377106 GHz, 377107.5 GHz, respectively, with the assumption of a noiseless and infinitely fast detector. Their scales are adjusted (green /20, black × 3 and red × 1) to fit in one plot. (b) The wave shape recorded (scale adjusted) with a 1-ns-rise-time detector and oscilloscope: black dashed curve, red solid curve.

*4.2. Pulse delay*

The delay is defined as the arrival time of the highest-intensity peak with respect to the peak arrival time without the cell in the path. Delay of 4.5 ns is observed when a 200 ps pulse passes through the 75mm long Rb 85 vapor cell at 100 C. The field amplitude at 4 ns away from the peak center is estimated to be $10^{-482}$ for a Gaussian pulse, thus the delayed pulse is not caused simply by a temporally local absorption. Many studies [6, 7, 12] reported pulse advance or delay using µs [6, 12] or even ms [7] pulses, with delay or advance equal to only a portion of the pulse duration.

We observe a linear-dispersion-induced pulse delay of 20 times the input pulse width. Such an effect has not been previously reported to our knowledge. Although there is no fundamental limit to the fractional time delay experienced by an optical pulse through a slow-light medium [22], the maximum fractional time delay reported to date appears to be the value of approximately 10 reported by H. Tanaka *et al*. [3]. We were not able to resolve the duration of the transmitted pulse, although the simulation in Figure.5 predicts it should be about 1.5 ns.

In all 'superluminal' experiments reported so far, the pulse advance is small compared with the pulse width [7, 12, 23]. That is limited by the quantum noise from the spontaneous emission and superfluorescence for the inverted system [23]. The advance must be shorter than the pulse duration to guarantee better signal to noise ratio (SNR). The pulse delay and advance in the Chu and Wong experiment [4] are less than half of the 50 ps pulse width since a 10 µm GaP:N crystal is used. The reason for choosing a short dimension is to minimize the absorption loss. Thus, a non-absorption dispersion model [24] was used to compare with the result.

In our experiment, the transmission of pulse energy with the frequency tuned between two absorption lines is about 5%. This is smaller than the 55% transmission in EIT experiments using a 10 cm Pb vapor cell [25, 26]. That is because our medium absorbs two wings of our broad input pulse spectrum and allows the center part to propagate with lower absorption. The slower the group velocity, the higher attenuation the pulse has. This is an intrinsic property of linear dispersion.

*4.3. Frequency dependence of pulse delay*

We attain good agreement with our theoretical prediction for pulse delay versus the pulse center frequency, shown in Figure. 6. As we tune the laser frequency toward the mid-point of the two absorption lines, the second temporal peak gets bigger and the first peak gets smaller and eventually disappears (Figure. 5. b). Near the two resonant regions, the two peaks are comparable in amplitude (Figure. 4. b, black trace). A small change in frequency in this region will make one peak bigger than the other peak. Thus, a sharp jump in the pulse delay is expected and observed. These jumps are also observed in the simulation in Figure. 6. There are some small discrepancies between experiment and simulation in the resonant regions. This probably arises from an observed $\pm 0.2$ GHz frequency jitter from the laser, making the wave shape unstable. The slope is negative in certain regions of the imaginary part of susceptibility, which implies "superluminal" propagation for narrow-linewidth pulses (also see Figure. 3 a.). Nevertheless, for both our experiment and simulation of broad-linewidth short-pulse propagation, the pulse is delayed in all regions. For the broad-linewidth input spectrum, some components fall in the positive slope parts of the refractive-index-versus-frequency curve, and are delayed. The other components fall in the negative slope part and should be advanced. However, the value of imaginary susceptibility in the negative slope part is large [3]. Thus, those components are highly absorbed. The overall pulse is observed to be delayed.

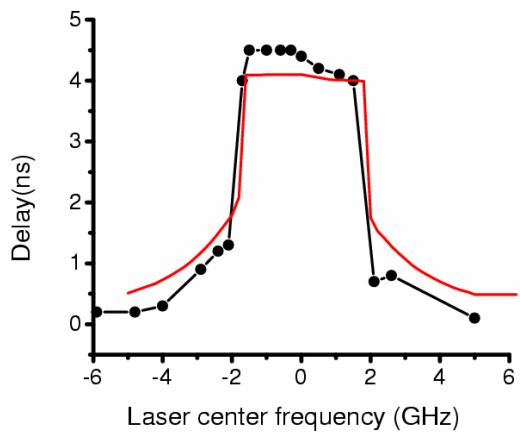

Figure. 6. The black curve with square points is the experimental observation of pulse delay vs. pulse center frequency. The red curve is the simulation of delay vs. frequency, using no free parameters.

*4. Control of delay by optical pumping*

We investigated the effect of optical pumping on the delay (Figure. 7). The ground-state atomic populations were be modified by optical pumping. By applying a CW diode laser beam, which is resonant with F=2 to F'=2, 3 transition, the atomic population accumulates on the F=3 level. If we assume 100% transfer efficiency, the resulting energy structure can be treated as a 2-level system. A Sharp laser diode (GH0781JA2) with an ILX LDX-3620 current driver and a Thorlab temperature controller is used for optical pumping. It runs continuous-wave (CW) with output power of 50 mW and spot size of 5 mm by 2 mm measured right before the cell. An AC (MHz) signal added to the driver modulation input can increase the optical linewidth of the pumping. A polarizer is used to keep the polarization of pumping light the same as that of the pulsed probe light. In order for the optical pumping to be effective, given the limited pumping power, the Rb gas temperature has to be reduced to 88 C with number density of $2.1 \times 10^{18}$ m$^{-3}$. In the F=3 to F'=2, 3 transition region, the delay is observed to equal 30% of the delay in the unpumped situation. The group velocity is faster in the optical-pumping situation and thus makes the pulse delay adjustable. Thus, two pulses could be synchronized by controlling the optical pumping intensity.

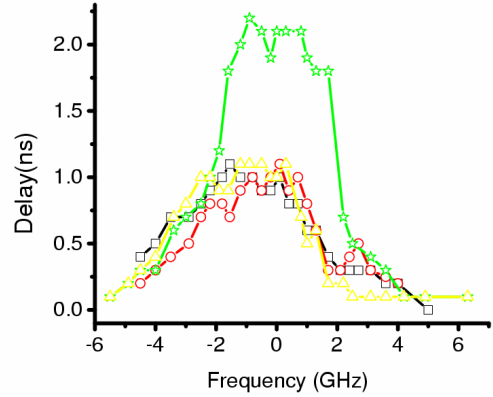

**Figure. 7.** Optical pumping effect on the delay. The delay is measured when pulse passes through the 75 mm long, 88 C Rb 85 vapor cell. The green (star) curve is the delay vs. frequency when there is no pump. The black (square), red (circle) and yellow (triangle) curve is the delay vs. frequency when Rb 85 is pumped by CW light with corresponding linewidth of 590 MHz (equal to Doppler linewidth), 1.1 GHz (greater than Doppler linewidth), 200 MHz (sub-Doppler linewidth).

*4.5. Discussion*

As we change the temperature gradually from room temperature up to 100 C, an increase of delay due to an increase of number density is observed, which is consistent with the theory. On the other hand, if we increase the probe pulse power, the delay becomes smaller. That is likely because more atoms are excited to a higher level, thus making the slope in the susceptibility smaller, and the group velocity approaches the vacuum light speed. In the optical pumping case, there is no noticeable difference in the plot of delay vs. center frequency for different pumping linewidth, because the optical pumping intensity of 500 mW/cm$^2$ is far beyond the saturation level of 75mm long Rb 85 vapor at 100 C [17]. In contrast, in self-induced transparency (SIT) experiments, the pulse has a long delay for all situations in which the input intensity is above a critical value. In our experiment, maximum delay occurs only at low input intensity. The transmission is also quite small at maximum delay compared with SIT experiment. This behavior rules out SIT-like effects in our experiment [11].

There are still some open questions. The measured pulse shape is somewhat narrower than that in the simulation (Figure. 5). This relation should be reversed, given the 1-ns rise and fall time of the detector. For modeling the optical pumping situation, we assume 100% population transfer and use a single ground level to calculate the delay which is around 1.4 ns. But the delay of 1 ns observed in the experiment (Figure. 7) is shorter than the simulated delay by 40%.

## 5. Gain doublet modeling

Given that our model is adequate for describing the main features of propagation through an absorption doublet, we should be able to apply our model to a gain-doublet medium. To explore this concept, we invent an energy level structure that is similar to Rb 85, but inverted. There are two ground states g and f, and two long-lifetime excited states $e_1$ and $e_2$, which are 3 GHz apart and approximately 377107 GHz above the ground states. By applying optical pumping, the population in state g and f is transferred to $e_1$ and $e_2$ with populations $\rho_{e1} = 5/12$, $\rho_{e2} = 7/12$. Thus, an incoherent gain-doublet structure is formed between the excited states and ground states. Right after the transfer process, a test pulse is sent in. We assume a low number density ($4 \times 10^{17}$ $m^{-3}$, which is around 10% of number density at 100 C) to guarantee that there is no population saturation during amplification. The other parameters are the same as in the absorption simulations above.

Because there are two frequencies that are predominantly amplified, oscillations appear in the pulse intensity. The oscillating structure of the wave shape makes the peak delay time change suddenly when we tune the test pulse frequency. In order to describe the pulse propagation behavior, we introduce an energy delay that is an intensity-weighted time delay.

$$\tau_D = \frac{\int_{-\infty}^{\infty} I(t)t\,dt}{\int_{-\infty}^{\infty} I(t)\,dt} \qquad (5)$$

From the simulation, we find that the transmitted pulses become temporally stretched more than in the absorbing situation. We also find pulse advance rather than pulse delay, with absolute values around five times larger than the delays found in the absorbing situation. The linear susceptibility and the delay/advance versus frequency are given in Figure. 8.

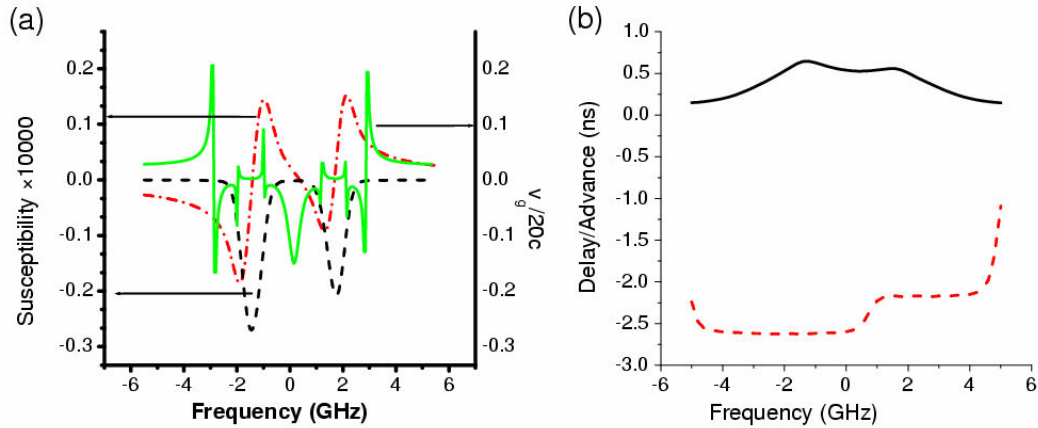

**Figure. 8**. Theoretical calculation of (a) the real part (black dashed, scale adjusted) and the imaginary part (red dash dotted, scale adjusted) of linear susceptibility, the group velocity (green solid) for near-monochromatic pulses, vs. laser frequency. The sharp changes in the group velocity vs. frequency near resonant regions are not observable in experiments using short pulses. (b) The energy delay/advance, defined in Eq. (5), for a 2.3 GHz linewidth pulse vs. frequency, for gain (red dashed) and absorbing situation (black solid). The positive/negative signs on the vertical axis indicate pulse energy delay/advance.

## 6. Conclusion

In summary, delay of 20 times the input pulse width, and group velocity $c/20$ are observed when a 200 ps pulse is incident on hot Rb 85 vapor. Measurements of pulse-peak delay confirmed that delay is caused by the linear response of Rb 85 vapor cell. The linear dispersion model for a three-level $\Lambda$ system yields good agreement with most of the experimental observations. We further demonstrated the controllability of delay by ground state optical pumping, and we explored the application of the theoretical model in a gain-doublet model system.

**Acknowledgment**

This work has been done under NSF support with grants number PHY-0140370 and PHY-0456974.